\documentstyle[prc,aps,preprint,tighten]{revtex}
\begin{document}

\draft

%\twocolumn [ \hsize\textwidth\columnwidth\hsize\csname
%@twocolumnfalse\endcsname

\title{Variational separable expansion scheme for two-body
Coulomb-scattering problems}
\author{J.\ Darai$^{1}$, B.\ Gyarmati$^{2}$, 
B.\ K\'onya$^{2}$ and Z.\ Papp$^{2,3}$}
\address{
$^1$ Institute of Experimental  Physics, University
of Debrecen,  Bem t\'er 18/a, H--4026 Debrecen, Hungary \\
$^2$ Institute of Nuclear Research of the Hungarian
Academy of Sciences,  
P.O. Box 51, H--4001 Debrecen, Hungary \\
$^3$ Department of Physics and Astronomy, 
California State University, Long Beach, CA 90840, USA 
}

\date{\today}

\maketitle

\begin{abstract}
We present a separable expansion approximation method for
Coulomb-like potentials which is based on Schwinger variational
principle and uses Coulomb-Sturmian functions as basis states.
The new scheme provides faster convergence with respect to our formerly
used non-variational approach. 
\end{abstract}
\vspace{0.5cm}

\narrowtext

%]

Both variational approaches and separable expansion schemes are
extensively used in solving few-body problems. Some time ago Adhikari
and Tomio presented an unified treatment of separable expansion schemes
based on Schwinger variational principles \cite{adhik}.
They proposed various approximation schemes
for the transition operator  $t$ satisfying the Lippmann-Schwinger  equation
\begin{equation}
\label{LS00}
t=v+vg^0t,
\end{equation}
where $v$ is the potential and $g^0$ is the free Green's operator. 
It was found that using these schemes 
with appropriate choice of expansion functions a rapid convergence could be 
obtained. However, in this paper, not a single word 
was devoted to Coulomb-like potentials.

At about the same time in a series of papers \cite{papp1} 
a separable expansion scheme for Coulomb-like potentials 
was proposed by one of us.
The Coulomb interaction was kept in the Green's operator and only
the short range part of the potential was subject to the separable expansion.
This approach uses  Coulomb-Sturmian (CS) functions as basis allowing
an exact analytical calculation of the matrix
elements of the Coulomb Green's operator
(see \cite{papp1}, and recently \cite{klp}). 
Thereby only the short-range part of the interaction 
is approximated and the correct Coulomb asymptotic properties 
of all quantities are guaranteed. 
The method has also been applied in three-body Faddeev calculations 
for bound-state 
and scattering problems with Coulomb interactions \cite{pzwp}.

In this paper we generalize one of the separable 
expansion schemes proposed by Adhikari and Tomio 
for two-body Coulomb-scattering problems. 

The expansion schemes in Ref.\ \cite{adhik} are based on a finite rank 
$N$ approximation of the product of operators
\begin{equation}
\label{ex1}
A B^{-1} C \approx \sum_{i,j}^N A|\eta_i \rangle D_{ij} \langle \zeta_j|C  ,
\end{equation}
where
\begin{equation}
\label{ex2}
(D^{-1})_{ji}= \langle \zeta_j|B| \eta_i \rangle.
\end{equation}
Here the sets  $| \eta_i \rangle $ and $| \zeta_j \rangle $ are assumed to be  
complete sets of states in the Hilbert space. If $N$ tends to 
infinity Eq.\ (\ref{ex1}) becomes an identity.
 
From among the separable approximation schemes of Ref.\ \cite{adhik}  
let us start with the form of $t$ following from 
Eq.\ (\ref{LS00})
\begin{equation}
\label{t0}
t={\bf 1}(v^{-1}-g_0)^{-1}{\bf 1}.
\end{equation}
Applying Eq. (\ref{ex1}) we get
\begin{equation}
\label{t01}
t \approx \sum_{i,j}^N |\eta_i \rangle D_{ij} \langle \eta_j| 
\end{equation}
with
\begin{equation}
\label{t02}
(D^{-1})_{ji}=\langle \eta_j| v^{-1}| \eta_i \rangle  
-\langle \eta_j|g_0| \eta_i \rangle.
\end{equation}
The first term in Eq.\ (\ref{t02}) can again be approximated as before
\begin{eqnarray}
\label{t03}
\langle \eta_j| v^{-1}| \eta_i \rangle & = &
\langle \eta_j|{\bf 1}v^{-1}{\bf 1}| \eta_i \rangle \nonumber \\
&\approx & \sum_{i',j'}^N \langle \eta_j|\zeta_{j'} \rangle 
C_{j' i'}
\langle \zeta_{i'}|\eta_i \rangle,
\end{eqnarray}
where
\begin{equation}
(C^{-1})_{i' j'}= \langle \zeta_{i'}|v| \zeta_{j'} \rangle .
\end{equation}

This approximation scheme can easily be generalized for Coulomb-like
potentials. A Coulomb-like potential in some 
partial wave $l$ can be written in the form
\begin {equation}
\label{potc}
v_l=v^C+v^s_{l},
\end{equation}
where $v^C$ is the pure Coulomb potential and $v^s_{l}$ is short ranged.
The Lippmann-Schwinger equation for the Coulomb-modified 
transition operator $t^{sC}$  reads
\begin{equation}
\label {LST}
t_{l}^{sC}=v_{l}^s+v_{l}^s g_l^C(E)t_{l}^{sC},
\end{equation}
where $g_l^C(E)=(E-h_l^0-v^C)^{-1}$ is the Coulomb Green's operator and 
$h_l^0$ is the free Hamiltonian. The solution of this equation can be given
in a form  analogous to Eq.\ (\ref{t0})
\begin{equation}
\label{transc}
t_{l}^{sC}={\bf 1}\left[(v^s_{l})^{-1}-g_l^C(E)\right]^{-1}{\bf 1},
\end{equation}
and the whole procedure of Eqs.\ (\ref{t01})-(\ref{t03}) can be repeated.
Only the free Green's operator has to be replaced by the Coulomb one.

As basis states we choose CS functions because they 
allow an exact analytical calculation of matrix elements of $g_l^C(E)$.
In coordinate representation they have the form
\begin{equation}
\langle r\vert{n;b} \rangle= \left[ \frac{n!}{(n+2l+1)!} 
\right]^{1/2}(2br)^{l+1}
\exp{(-br)}L_n^{2l+1}(2br),
\end{equation} 
where $n=0,1,2,...$, $L_n^{2l+1}$ are the Laguerre polynomials and $b$ is a 
scaling parameter. 
They have an analogous simple form also in momentum representation
\begin{eqnarray}
\langle p|n;b\rangle 
 &=& \frac{2^{l+3/2}l!(n+l+1)\sqrt{n!}}{
\sqrt{\pi (n+2l+1)!}}  \nonumber \\
&&\times \; \frac{b (2bp)^{l+1}}{(p^2+b^2)^{2l+2}}G_n^{l+1}
\left(\frac{p^2-b^2}{p^2+b^2}
\right) ,  \label{basisp}
\end{eqnarray}
where $G$ denotes the Gegenbauer polynomials.
Introducing the notation $\langle r\vert \widetilde{ n;b} \rangle =
\langle r|{n;b} \rangle/r$ we can express the orthogonality and completeness of
the CS functions as 
\begin{equation}
\langle {n;b} \vert \widetilde{ n^{\prime};b} \rangle =\delta_{nn^{\prime}}
\label{csort}
\end{equation}
and 
\begin{equation}
\label{unit}
{\bf 1}  =  \lim_{N\to\infty}\sum _{n=0}^N 
\vert \widetilde{n;b} \rangle \langle {n;b} \vert 
 =  \lim_{N\to\infty}\sum_{n=0}^\infty \vert {n;b} \rangle 
\langle \widetilde{n;b} \vert,
\end{equation}
respectively. 
 
As $| \eta_i \rangle $ and $| \zeta_j \rangle $ we take
CS bases with different $b$ parameters:
\begin{equation}
\label{t1}
t_{l}^{sC} \approx 
\sum_{n,n^{\prime}=0}^N \vert \widetilde{n;b_1} \rangle
D_{nn^{\prime}} \langle \widetilde{ n^{\prime};b_1} \vert,
\end{equation}
\begin{eqnarray}
\label{t2}
(D^{-1})_{n^{\prime}n} & = & \sum_{m,m^{\prime}=0}^N\langle 
{n^{\prime};b_1}\vert\widetilde{m;b_2} \rangle 
C_{m m'}  \langle\widetilde{m^{\prime} ;b_2}
\vert {n ;b_1}\rangle \nonumber \\
&&  - \langle {n^{\prime};b_1}\vert g_l^C \vert {n;b_1}\rangle,
\end{eqnarray}
\begin{equation}
\label{t3}
(C^{-1})_{m' m}=  \langle {m^{\prime};b_2}\vert v^s_{l}\vert
 m ;b_2 \rangle .
\end{equation}
While the matrix elements of the potential and the overlap of CS functions
should be calculated numerically either in configuration or momentum space
the matrix elements of the Coulomb Green's operator
can be calculated analytically  \cite{papp1,klp}.

This scheme is essentially equivalent with a twofold separable expansion 
for potential  $v^s_{l}$
\begin{eqnarray}
\label{potapp}
 v^s_{l}  \approx  
\sum_{n,n^{\prime},m,m^{\prime}=0}^N  &&
\vert \widetilde{n;b_1} \rangle (\langle{n;b_1} \vert
\widetilde{m^{\prime};b_2}\rangle)^{-1} \nonumber \\
&& \langle{m^{\prime};b_2}\vert v^s_{l} \vert {m ;b_2}\rangle \nonumber \\
&& ( \langle \widetilde{ m ;b_2}\vert n^{\prime};b_1 \rangle)^{-1} 
\langle \widetilde{ n^{\prime} ;b_1} \vert .
\end{eqnarray}
It is easy to verify that
Eqs. (\ref{t1})-(\ref{t3}) are the solutions of the Lippmann-Schwinger equation Eq. 
(\ref{LST}), with this approximate potential.

This scheme should be compared with the 
CS potential separable expansion scheme 
of Refs.\ \cite{papp1,klp} which is not supported by variational principle.
It is based on the approximation of the unit operator
\begin{equation}
 {\bf 1}  = 
\lim_{N\to\infty} \sum_{n=0}^N \vert\widetilde{n}\rangle \sigma_n^N
\langle {n}\vert,
\end{equation}
with $\sigma$ factors possessing the properties $\lim_{n\to\infty} \sigma_n^N =0$
and $\lim_{N\to\infty} \sigma_n^N =1$.
Now, the approximation takes the form
\begin{equation}
\label{orig}
v^s_{l} ={\bf 1} v^s_{l}{\bf 1} \approx  \sum_{n,n^{\prime}=0}^N 
\vert\widetilde{n}\rangle \sigma_n^N
\langle {n}\vert v_{sl} \vert n^{\prime}\rangle 
\sigma_{{n^{\prime}}}^N \langle 
\widetilde{n^{\prime}}\vert .
\end{equation}
For $\sigma_n^N$ the form
\begin{equation}
\label{sigma}
\sigma_n^N = \frac{1-\exp\{-[\alpha(n-N-1)/(N+1)]^2\}}{1-\exp(-\alpha^2)}
\end{equation}
was used with $\alpha \sim 5$. This value of the arbitrary parameter
$\alpha$ yielded the fastest convergence.

To show the relative power of these separable expansion schemes we 
have calculated the Coulomb-modified nuclear phase shifts  $\delta^{sC}_l$ 
at $l=0$ and at various energies for a $p-p$ scattering. The short range
potential was taken in  Malfliet-Tjon form
\begin{equation}
\label{mf}
v^s_{l}= v_0 \exp(- \beta_0 r)/r +v_1 \exp(- \beta_1 r)/r,
\end{equation}
with $v_0=-626.885\mbox{MeV}$, 
$\beta_0=1.55\mbox{fm}$, $v_1=1438.720\mbox{MeV}$, $\beta_1=3.11\mbox{fm}$.
In the method of Eq.\ (\ref{orig}) CS basis parameter $b=3\mbox{fm}^{-1}$
was used, while in the expansion Eq.\ (\ref{potapp})  
$b_1=3.8\mbox{fm}^{-1}$ and $b_2=2.5\mbox{fm}^{-1}$ were taken.
 It can be seen in Table I that in both
approximation schemes it is possible to choose the $b$ parameters so that
the expansions give almost equally fast convergence over the whole spectrum and 
provide extremely accurate results. 
To reach 6-digits accuracy the method of Eq.\ (\ref{potapp}) 
needs 10-13 basis states,
while the method of Eq.\ (\ref{orig}) needs 20-23 states. 
We have observed similar results over a wide
range of $b$ parameters.
It should be noted, however, that the method of Eq.\ (\ref{potapp}) 
is more complicated numerically, 
so the numerical effectivity of both methods are
more or less the same. 

In this Brief Report we have combined two separable expansion 
methods. One based on Schwinger variational principle
has been  proposed by Adhikari and Tomio in Ref.\ \cite{adhik}.
The other approach proposed by Papp in Refs.\ \cite{papp1} is not variational
and it was designed for Coulomb-like potentials.
This new scheme is  a variational separable expansion method and
is applicable for Coulomb-like potentials. It converges considerably
faster in terms of basis states then the non-variational method
of Eq.\ (\ref{orig}).
This property could be useful in three-body calculations
where the rank of the expansion is of crucial importance. 
The method of Eq.\ (\ref{orig}) has been
generalized for solving Faddeev-type integral equations of three-body 
Coulombic systems. Whether or not the method of Eq.\ (\ref{potapp}) can 
be extended in this direction is still an open question.

This work has been supported by the OTKA Gratns No.\ T026233 and
T029003 and the NSF Grant No.Phy-0088936.

\widetext

\begin{table}
\begin{center}
\begin{tabular}{rcccccccc}
$N$ & \multicolumn{2}{c}{$E=0.1$ MeV } & \multicolumn{2}{c}{$E=1$ MeV}  
& \multicolumn{2}{c}{$E=10$ MeV}& \multicolumn{2}{c}{$E=100$ MeV}  \\
   & Eq.\ (\ref{orig})& Eq.\ (\ref{potapp}) &  Eq.\ (\ref{orig}) & 
   Eq.\ (\ref{potapp}) &  Eq.\ (\ref{orig})& Eq.\ (\ref{potapp}) & 
   Eq.\ (\ref{orig}) & Eq.\ (\ref{potapp})  \\ \hline       
 2  &  -0.100682 & -0.151364  &  -0.624200 & -0.797903   & 1.546662 & 
 1.453638 &  0.163942  &  0.253576  \\
 3  &  -0.115923 & -0.122391  &  -0.685037 & -0.710521   & 1.506720 & 
 1.479076 &  0.365679  &  0.351795  \\
 4  &  -0.120857 & -0.120468  &  -0.706248 & -0.705228   & 1.480305 & 
 1.477592 &  0.430684  &  0.403522  \\
 5  &  -0.118844 & -0.119136  &  -0.701335 & -0.701377   & 1.473793 & 
 1.477278 &  0.415215  &  0.402637  \\
 6  &  -0.118655 & -0.119182  &  -0.699997 & -0.701573   & 1.481240 & 
 1.479584 &  0.411720  &  0.404552  \\
 7  &  -0.119150 & -0.119165  &  -0.701393 & -0.701565   & 1.483505 & 
 1.480309 &  0.408933  &  0.407128  \\
 8  &  -0.119166 & -0.119165  &  -0.701534 & -0.701549   & 1.481915 & 
 1.480743 &  0.408615  &  0.407224  \\
 9  &  -0.119147 & -0.119164  &  -0.701390 & -0.701523   & 1.481590 & 
 1.480921 &  0.409217  &  0.407345  \\
10  &  -0.119155 & -0.119164  &  -0.701439 & -0.701519   & 1.481230 &
 1.480945 &  0.408458  &  0.407467  \\
11  &  -0.119157 & -0.119162  &  -0.701462 & -0.701508   & 1.481028 & 
1.480958 &  0.407939  &  0.407487  \\
12  &  -0.119158 & -0.119162  &  -0.701472 & -0.701507   & 1.480981 & 
1.480957 &  0.407830  &  0.407488  \\
13  &  -0.119160 & -0.119162  &  -0.701490 & -0.701505   & 1.480959 & 
1.480957 &  0.407694  &  0.407497  \\
14  &  -0.119160 & -0.119162  &  -0.701492 & -0.701504   & 1.480958 & 
1.480957 &  0.407585  &  0.407499  \\
15  &  -0.119161 & -0.119162  &  -0.701498 & -0.701504   & 1.480959 & 
1.480957 &  0.407557  &  0.407499  \\
16  &  -0.119161 & -0.119162  &  -0.701499 & -0.701504   & 1.480959 &
 1.480957 &  0.407535  &  0.407499  \\
17  &  -0.119161 & -0.119162  &  -0.701501 & -0.701504   & 1.480959 & 
1.480957 &  0.407511  &  0.407499  \\
18  &  -0.119161 & -0.119162  &  -0.701502 & -0.701504   & 1.480959 & 
1.480957 &  0.407506  &  0.407499  \\
19  &  -0.119161 & -0.119162  &  -0.701502 & -0.701504   & 1.480958 & 
1.480957 &  0.407506  &  0.407499  \\
20  &  -0.119162 & -0.119162  &  -0.701503 & -0.701504   & 1.480957 & 
1.480957 &  0.407500  &  0.407499  \\
21  &  -0.119162 & -0.119162  &  -0.701503 & -0.701504   & 1.480957 & 
1.480957 &  0.407499  &  0.407499  \\
22  &  -0.119162 & -0.119162  &  -0.701503 & -0.701504   & 1.480957 & 
1.480957 &  0.407501  &  0.407499  \\
23  &  -0.119162 & -0.119162  &  -0.701504 & -0.701504   & 1.480957 & 
1.480957 &  0.407499  &  0.407499  \\
24  &  -0.119162 & -0.119162  &  -0.701504 & -0.701504   & 1.480957 & 
1.480957 &  0.407499  &  0.407499  \\
25  &  -0.119162 & -0.119162  &  -0.701504 & -0.701504   & 1.480957 & 
1.480957 &  0.407499  &  0.407499  \\
26  &  -0.119162 & -0.119162  &  -0.701504 & -0.701504   & 1.480957 & 
1.480957 &  0.407499  &  0.407499  \\
27  &  -0.119162 & -0.119162  &  -0.701504 & -0.701504   & 1.480957 & 
1.480957 &  0.407499  &  0.407499  \\
28  &  -0.119162 & -0.119162  &  -0.701504 & -0.701504   & 1.480957 & 
1.480957 &  0.407499  &  0.407499  \\
29  &  -0.119162 & -0.119162  &  -0.701504 & -0.701504   & 1.480957 & 
1.480957 &  0.407499  &  0.407499  \\
\end{tabular}
\end{center}
\caption{
Convergence of the Coulomb-modified nuclear phase shift 
$\delta^{sC}_0(E)$ (in radians) 
in the (\ref{mf}) potential at different energies with respect to the 
$N$ number of basis states used in the separable expansion schemes
(\ref{orig}) and  (\ref{potapp}).
\label{tab:method_a}}
\end{table}

\end{document}